\documentclass[aps,prb,twocolumn,showpacs,floatfix,superscriptaddress,amssymb,
amsmath]{revtex4-1}
\usepackage{graphicx}
\usepackage{dcolumn}
\usepackage{bm}
\usepackage{amsmath}
\usepackage{fontenc}
\newcommand{\dlangle}{\langle\langle}
\newcommand{\drangle}{\rangle\rangle}

\usepackage{color}

\begin{document}
\title{Suppression of Kondo screening by the Dicke effect in multiple quantum dots}
\author{E. Vernek}
\affiliation{Instituto de F\'isica - Universidade Federal de Uberl\^andia,
Uberl\^andia, MG  38400-902, Brazil}
\author{P. A. Orellana}
\affiliation{Department of Physics, Universidad Cat\'olica del Norte, Casilla 1280, Antofagasta, Chile}
\author{S. E. Ulloa}
\affiliation{Department of Physics and Astronomy, and Nanoscale
and Quantum Phenomena Institute, \\Ohio University, Athens, Ohio
45701-2979}

\date{\today}
\begin{abstract}
The interplay between the coupling of an interacting quantum dot to a
conduction band and its connection to localized levels  has been studied in a triple quantum dot
arrangement. The electronic Dicke effect, resulting from quasi-resonant states of
two side-coupled non-interacting quantum dots, is found to produce important effects
on the Kondo resonance of the interacting dot. We study in detail the Kondo regime of the system
by applying a numerical
renormalization group analysis to a finite-$U$  multi-impurity Anderson Hamiltonian model.
We find an extreme narrowing of the Kondo resonance, as the single-particle levels of
the side dots are tuned towards the Fermi level and ``squeeze" the Kondo resonance, accompanied
by a strong drop in the Kondo temperature, due to the presence of a supertunneling state.
Further, we show that the Kondo temperature {\em vanishes}
in the limit of the Dicke effect of the structure.   By analyzing the
magnetic moment and entropy of the three-dot cluster versus temperature, we identify a
different {\em local} singlet that competes with the Kondo state, resulting in the eventual
suppression of the Kondo temperature and  strongly affecting the spin correlations of the structure.
We further show that system asymmetries in couplings, level structure or due to Coulomb interactions,
result in interesting changes in the spectral function near the Fermi level.  These strongly affect the 
Kondo temperature and the linear conductance of the system.

\end{abstract}

\maketitle

\section{Introduction}

Quantum dots (QDs) have played a prominent role in the investigation of the Kondo
problem physics in recent years, \cite{DGG,Inoshita,Sasaki,Cronenwett,Pustilnik}
as they allow a systematic and well-controlled variation of structure parameters, with
the consequent exploration of electronic correlations in this paramount many-body state.
This exploration is of course possible by being able to
experimentally tune the relevant parameters in the system
over rather wide ranges.
Moreover, in a feature that is unique to these systems, QDs also
facilitate the incorporation and study of quantum coherence effects in the
Kondo state, and its interplay with structure resonances and field-induced phase shifts.
In fact, the Kondo effect has been studied in multi-quantum dot  systems, including
double,\cite{busser,jeong,cornaglia,vernek,8l,sasaki3,zitko5} and
triple dots, \cite{kuzmenko1,jiang,zitko2,lobos,Vernek1,Numata,Chiappe1,klaus,haug} in
different geometrical arrangements, motivated in part by the richness of the Kondo
physics in the presence of localized levels, as well as by their
potential application as spin filters.

One of the main signatures of
the Kondo state is the enhancement of the quasi-particle density of states at the Fermi
level; this Kondo resonance is accessible in
electronic transport experiments, as it opens an additional transport channel, which
is readily seen in the differential conductance of these structures (typically in the zero-bias
limit).
In the case of multiple QD geometries, the
Kondo and other single-particle--like resonances provide different transport channels which
can even interfere with one another.  As a
result, these structures offer the interesting possibility of controlling the
transport properties by exploiting quantum interference in the electronic propagation, allowing
the study of scattering phenomena such as the
well-known Fano \cite{fano} and Aharonov-Bohm effects,\cite{Hofstetter,17'} competing with the Kondo
 effect.
For example, the interference of  Kondo and  Fano resonances in suitably designed structures has
been  shown recently
in beautiful experiments, giving rise to complex conductance features.\cite{sato,sasaki3}
These results demonstrate that consideration of multiple coherent scattering of traveling
electronic waves, both through single-particle as well as many-body resonances, is crucial for the
understanding of the resulting conductance features.  In this regard, being able to tune the
structure  parameters
of a multiple QD system, as well as its Kondo state, provides a unique arena to study correlations
and coherence
effects in a controllable manner.

It is clear that the features of the Kondo resonance near the Fermi level significantly affect
the conductance of the system, sometimes in quite subtle ways.  For example, the
shape of the density of states in the leads near the FL has been shown
to produce strong modifications of the Kondo resonance and characteristic energy (the Kondo
temperature).   These
modifications can arise from intrinsic global properties,\cite{WF,Ingersent} or from the local
environment
in the vicinity of the active QD.\cite{8l}
We are here especially interested in resonances and modifications due to the
electronic version of the Dicke effect. \cite{orellana}   The latter is the electronic analogue
of the well-known Dicke effect in quantum optics, which takes place in
the spontaneous emission of closely-linked atoms lying in the same environment (within one
characteristic
wavelength of each other).\cite{dicke} In the electronic
case, the decay rates (level broadenings) are produced by
the couplings between localized levels and a conduction channel, and their close proximity and
effective coupling gives rise to effectively fast
({\em super-tunneling}) and slow ({\em sub-tunneling}) modes.\cite{18S,18C,Brandes}  Interestingly,
this coherent single-particle
physics results in strong changes of the Kondo screening, as we will discuss below, once Coulomb
interactions are fully taken into account.

Although other multidot geometries would exhibit similar physics, a specific configuration to study
this effect consists of three QDs where two large (essentially non-interacting)
dots are attached laterally to a central dot that is embedded between current leads, as
schematically shown in Fig.\ \ref{fig1}. This cross-bar
structure is reminiscent of quantum wave guides with a resonator cavity.
\cite{Debray} In our case, the side QDs act as scattering centers to the main transport channel,
and compete with the Kondo effect of the central dot.\cite{torio}
Similar configurations have been studied before, and recent work by Trocha and Barnas using a slave
boson
mean field approach has identified interesting regimes.\cite{Trocha}
They studied the  interplay between Kondo and Dicke
resonances and considered the limit of infinite Coulomb repulsion ($U\rightarrow
\infty$), which suppresses virtual processes  involving the  doubly occupied state in the
interacting dot.  Additionally,  the mean-field approximation adopted in the auxiliary boson fields
neglects correlations due to charge fluctuations involving the empty state of the dot.  All of these
processes may be expected to be important
in providing a full description of the Kondo resonance, especially in competition
with the Dicke effect here, as we indeed show in this work.  Our numerical renormalization
group (NRG) approach, \cite{Wilson,Murthy,24B} provides a reliable description that incorporates all
charge  and spin fluctuations in the problem, it allows the study of finite $U$
values--corresponding to real QD experiments--and gives us
a systematic way to study the subtle interference effects inherent in this geometry.

In this work, we explore the Kondo state
that appears when the levels of the non-interacting dots are brought together both
symmetrically and asymmetrically about  the Fermi level. In both situations, the Kondo resonance
narrows drastically and is ``squeezed'' between the two single-particle levels,
suggesting a drop in the Kondo temperature of the system, as described recently in the symmetric case.\cite{Trocha}
We show that indeed the Kondo temperature drops, as the onset of the Dicke effect involving the interacting dot
totally suppresses the Kondo state and results in {\em vanishing} Kondo temperature.
This behavior, due to the enhancement of spin-spin correlations in the QDs orbitals, results in the
formation
of a {\it local} singlet, which decouples from the current leads. As the system moves
away from a fully degenerate Dicke configuration, the Kondo temperature is non-zero but has a
non-monotonic
dependence on structure parameters, behavior which is only brought out by our numerical
renormalization group approach. This optimization of the Kondo effect in a structure may have
interesting
applications in a real experimental system.

We will further show that the competition of Kondo and Dicke effects modifies
excited states near the Kondo regime as well, resulting in unexpected changes in the thermodynamic
and transport properties of the system.  Interestingly, the local (non-Kondo) singlet state
mentioned above emerges due to a strong
coupling between  the interacting QD and the super-tunneling state. A
crossover between these two configurations can be tuned by varying the coupling between the QDs.

We also describe the effects of asymmetries in the system, including different interdot couplings, as well as energy levels and 
finite Coulomb interactions in all the dots of the structure.  We show that these structural asymmetries introduce interesting particle-hole asymmetries in the problem, resulting in strong changes in the Kondo temperatures and the linear conductance of the system.  

\section{Model and theoretical approach} \label{model}
\begin{figure}
\centerline{\resizebox{2.5in}{!}{
\includegraphics{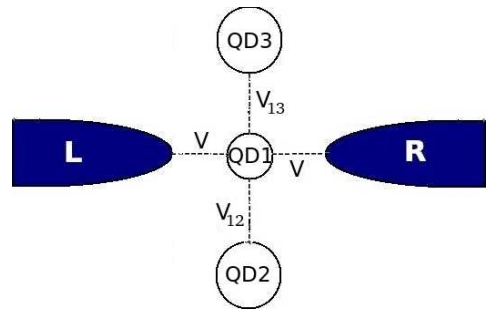}}}
\caption{\label{fig1} (color online) Schematic representation of the three-dot system.  QD1 is
considered to be  strongly interacting ($U\neq 0$), whereas QD2 and QD3 are assumed to be
effectively non-interacting ($U'$ small or vanishing--see text). }
\end{figure}

The system consists of an interacting QD (labeled QD1) coupled to two current leads
($L$ and $R$) and  to two effectively non-interacting QDs (QD2 and QD3) (see
Fig.\ \ref{fig1}); one could also think of QD2 and QD3 as being close to a Coulomb blockade peak, so
that they conduct and their behavior is single-resonance--like, as charging effects are typically
small when conducting.  The three-dot structure is described by a multi-impurity Anderson
Hamiltonian, $H=H_{dots}+H_{leads}+H_T$,
with
\begin{eqnarray}\label{H_dots}
H_{dots}&=&\sum_{\sigma, i=1}^3\varepsilon_ic^\dagger_{i\sigma}c_{i\sigma}
+U_i n_{
i\uparrow}n_{i\downarrow}+
\nonumber\\
&&+ \sum_{\sigma}\left[V_{12} c^\dagger_{1\sigma} c_{2\sigma}+ V_{13} c^\dagger_{1\sigma} c_{3\sigma}
+h.c.\right], \\
 H_{leads}&=&\sum_{{ k\sigma} ,\ell=R,L} \varepsilon_{\ell k}c^\dagger_{\ell
k\sigma}
c_{\ell k\sigma} \\
 H_T&=&V\sum_{\ell k\sigma \atop
\sigma}\left[c^\dagger_{1\sigma}c_{\ell
k\sigma}+h.c.\right],
\end{eqnarray}
where $c^\dagger_{i\sigma}$ ($c_{1\sigma}$) is the
operator that creates (annihilates) an electron with  energy $\varepsilon_i$
($i=1,2,3$) and spin $\sigma$ in the  respective QD and $c^\dagger_{\ell
k\sigma}$  is the corresponding fermion operator for
the leads $\ell=R,L$, with energy $\varepsilon_{\ell k}$, where $k$ is the
momentum
quantum number of the free conduction electrons. The second term in
Eq.\ (\ref{H_dots}) accounts for the Coulomb repulsion in the doubly
occupied state of the dots. The hopping amplitudes $V$, $V_{12}$ and $V_{13}$
couple the interacting QD1 to the leads and to QD2 and QD3,
respectively.  For simplicity, unless stated otherwise (see Sec.\ \ref{phasym}), we will take
$V_{12}=V_{13}= V^\prime$, and $U_1=U\neq 0$, while $U_2=U_3=0$.

The characteristic properties of the Kondo physics of
the system will be probed through thermodynamic
quantities, namely the local magnetic moment and its contribution to the entropy, which can be
readily obtained from the NRG procedure. To calculate dynamical quantities, such as
the spectral functions and the conductance of the system, one needs to calculate
the local retarded  Green's function (GF) at the interacting
dot, which is defined in the standard form,
\begin{eqnarray}
G_{11}^\sigma(\omega)&\equiv& \dlangle
c_{1\sigma};c^\dagger_{1\sigma}\drangle_\omega\nonumber\\
&=&\int_{-\infty}^\infty e^{-i\omega\tau}\dlangle
c_{1\sigma}(t);c^\dagger_{1\sigma}(t^\prime)\drangle d\tau,
\end{eqnarray}
where $\tau=t-t^\prime$, and
\begin{eqnarray}
\dlangle c_{1\sigma}(t);c^\dagger_{1\sigma}(t^\prime)\drangle=
-i\Theta(t-t^\prime)\langle
[c_{1\sigma}(t),c^\dagger_{1\sigma}(t^\prime)]_+\rangle
\end{eqnarray}
is the double-time Green's function, with $\Theta(\tau)$ the usual step function,
and $[\cdots,\cdots]_+$  the anticommutator. Once
$G_{11}^\sigma(\omega)$ is known, the spectral function $\rho_{1\sigma}(\omega)$ [or
density of states (DOS)] of the interacting dot can be obtained from the relation
\begin{eqnarray}\label{rho1eq}
\rho_{1\sigma}(\omega)=-\frac{1}{\pi}{\tt Im}[G_{11}^\sigma(\omega)].
\end{eqnarray}

\subsection{Non-interacting analysis}

In the non-interacting limit ($U_i\rightarrow
0$) the GF  $G_{11}^{(0)\sigma}(\omega)$
can be  easily  obtained by the equation-of-motion method, leading to an exact
closed expression,
 \begin{eqnarray}\label{Green0}
 G^{(0)\sigma}_{11}(\omega)=\frac{1}{\omega-\varepsilon_1-\Sigma_{1}(\omega)},
 \end{eqnarray}
where
\begin{eqnarray}
\Sigma_{1}(\omega)=\Lambda_{1}(\omega)+i\Delta_1(\omega) \label{eqS}
\end{eqnarray}
is the self-energy in the non-interacting case, which takes into account the
effects of the QD1 contacts with the leads, and with QD2 and QD3. The
self-energy can be written as
\begin{eqnarray}\label{sigma}
\Sigma_{1}(\omega)=2V^2\sum_k\frac{1}{\omega-\varepsilon_k}+V^{\prime
2}\frac{2\omega-\varepsilon_2-\varepsilon_3}{
(\omega-\varepsilon_2)(\omega-\varepsilon_3)}.
\end{eqnarray}
 In these expressions, the frequency $\omega$ must be understood in its
analytical  continuation sense, $\omega\rightarrow \omega+i\eta$, where $\eta$
is an infinitesimal. Notice that the poles at $\varepsilon_2$ and $\varepsilon_3$
appearing in
the self-energy describe localized states in the effective
conduction band due to the presence of QD2 and QD3. In the limit of
$V^\prime\rightarrow 0$, the self-energy
reduces to the single impurity case,
\begin{eqnarray}
\Sigma_{1}^{(V'=0)}(\omega)=\frac{\Delta_0}{2\pi}\ln \left| {\frac{\omega+D}{\omega-D}} \right|
-i\Delta_0,
\end{eqnarray}
where $\Delta_0=2\pi V^2/2D$ is the hybridization function for a flat  density of states,
and $D$ is the half-bandwidth of the
conduction electrons.  Notice that the effect of the
leads and the two side-coupled QDs is fully taken into account  in Eq.~(\ref{Green0}) by the
self-energy
$\Sigma_1(\omega)$, which provides a ``structured" effective conduction
band.  The electrons localized in QD1 are then coupled via the effective density of states in
the leads represented by $\Delta_1$ in Eq.\ (\ref{eqS}).

\begin{figure}
\centerline{\resizebox{3.3in}{!}{
\includegraphics[angle=0]{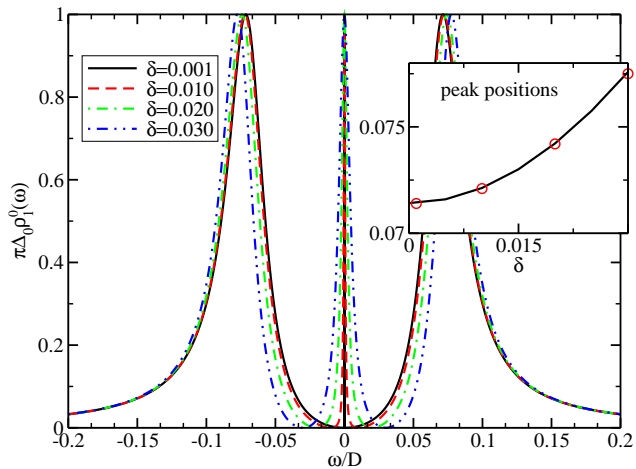}}}
\caption{\label{fig12}(color online)  Spectral function vs.\ energy for
$V=0.1$, $V^\prime=0.05$ and  $\epsilon_1=0$, for $\delta=0.001$ (solid, black), $\delta=0.01$
(dashed, red),  $\delta=0.02$ (dot-dash, green), and $\delta=0.03$ (dot-dot-dash,
blue). The inset shows  the position of the right-side satellite peak as function of $\delta$.
Solid line corresponds to the analytical expression (see text) while (red) circles show peak positions
taken from curves in main panel.}
\end{figure}

The spectral function, $\rho_1^0\equiv\rho_{\uparrow}=\rho_{\downarrow}$ in this
{\em non-interacting case} is depicted in Fig.\ \ref{fig12} for $\epsilon_1=0$, $V^{\prime}=0.05$, $V=0.1$
and different values of $\delta$. We observe that by increasing $\delta$ the width of the central peak
increases, while the satellite peaks are just slightly shifted away from the Fermi level. The shift
of the satellite peak position can be estimated by analyzing the poles of $G_{11}^\sigma$. Assuming
the satellite peaks to be Lorentzian-shaped, we find that their positions are given by
$\pm\sqrt{2V^{\prime 2}+ \delta^2}$. In the inset we show the peak positions as function of
$\delta$. The continuous (black) curve corresponds to the analytical expression $1.01\sqrt{2V^{\prime
2}+\delta^2}$, while symbols correspond to the values obtained
from the curves in the main panel. The need for the factor 1.01 results from the assumption that
the peak has a Lorentzian shape in the analytical estimate. On the other hand, close to zero energy ($\omega \sim 0$) and for
small detuning ($\delta \ll \Delta_0$),
the DOS can be written as a superposition of a symmetric Fano lineshape--arising from the super-tunneling Dicke state--of width $\Gamma_+\propto \Delta_0$, and a Lorentzian line--arising from the subtunneling state--
of width $\Gamma_{-}\propto \delta^2/\Delta_0$, respectively.\cite{orellana}
This complex behavior of the DOS is due to the hybridization of the lateral dots through the continuum of the central dot and leads. This phenomenon is in close analogy to the Dicke effect in quantum optics.\cite{orellana} Here, the super-tunneling and subtunneling modes give rise a broad ($\Gamma_+$) antiresonance and a sharp ($\Gamma_-$) resonance, respectively. As we will discuss later, this effect also modifies dramatically the many-body Kondo state.

\subsection{Interacting central dot}
We will now consider the interesting case of interactions in the central dot only, so
that $U_1=U$ is finite, while $U_2=U_3=U'=0$.  In this case, the system can
also be mapped into a single QD coupled to an effective conduction
band. In this case the dressed GF of the interacting dot can be formally
written as:
\begin{eqnarray}
G_{11}^\sigma(\omega)=\frac{1}{\omega-\varepsilon_1-\Sigma^*_{1}(\omega)},
\end{eqnarray}
where $\Sigma^*_1(\omega)$ is the proper self-energy which takes into account
the influence of the  lateral QDs, the current leads, as well as the Coulomb repulsion
$U$. Unlike the non-interacting case, a closed  expression for $\Sigma^*_1(\omega)$ is not
possible, due to the many-body term in the
Hamiltonian.  However, $G_{11}^\sigma(\omega)$ can be reliably obtained using the
NRG procedure.\cite{24B,Jayaprakash}  For convenience, we write  the GF in
the Lehmann representation,\cite{Fetter}
\begin{eqnarray}\label{Lehmann}
G_{11}^\sigma(\omega)=\frac{1}{Z}\sum_{nn^\prime}\frac{\mid \langle n|
c^\dagger_{d\sigma}|  n^\prime\rangle\mid^2}{\omega-E_{n^\prime}+E_n}\left(
e^{-\beta E_n}+e^{-\beta E_{n^\prime}}\right),
\end{eqnarray}
where $Z$ is the partition function in the canonical ensemble, $|n\rangle$
is an eigenstate of the
Hamiltonian $H$,  with corresponding  eigenenergy $E_n$, and $\beta=(k_BT)^{-1}$.
The imaginary part of $G_{11}^\sigma(\omega)$ is calculated
at zero temperature ($T=0$) directly from the NRG spectrum, \cite{Bulla1} while the real part
can be obtained from a Kramers-Kr\"onig transformation.\cite{Kramers}

\subsection{Cluster impurity approach} \label{CIA}

As we will see in the next section, it is also useful to analyze the details of the states involving the three QDs, considering
the system as a whole.  This approach also proves essential when dealing with Coulomb interactions in all three QDs, as
the approach of an effective density of states described in the previous section, is no longer applicable in the more general
case.  To that end, we
solve the problem considering all the dots composing the ``impurity region,"
so that the three-dot cluster  is coupled to a flat conduction
band.  This requires a simple generalization of the NRG procedure: start with a cluster impurity described by a set of 64 basis states,
on which one writes and diagonalizes the initial Hamiltonian, $H_{dots}$, while the current lead
is characterized by a flat density of states.  For these NRG ``cluster impurity" runs, we typically keep 1600 states and use a discretization parameter $\Lambda=3.5$, which provides sufficient accuracy for the spectral function in the scales of interest.

\section{The role of electronic interactions}
\subsection{Particle-hole symmetric case} \label{symmetric}
Hereafter, we set  $D=1$ (typically the largest
energy scale of the problem) as our energy unit.  We further
consider  the Coulomb repulsion $U=0.5$, and $\varepsilon_1=-U/2$, which
corresponds to a particle-hole symmetric system with interactions in QD1.
In order to study the Dicke effect in this system we first focus on the  particle-hole
symmetric point {\em of the entire system.} 
To this end, we tune the on-site energies  $\varepsilon_2$
and $\varepsilon_3$ of the non-interacting QDs ($U'=0$) symmetrically displaced from the
Fermi level, for which we set $\varepsilon_3=-\varepsilon_2=\delta$.
With this choice of parameters, the last term of Eq.\ (\ref{sigma}) shows two poles located at
$\omega=\pm\delta$, which  produce strong enhancement of the
effective coupling $\Delta_1(\omega)$. This results in a dramatic distortion of the local
interacting DOS projected in QD1, $\rho_{1}^\sigma (\omega)$, with important implications on the
 properties of the system. 
\begin{figure}[h]
\centerline{\resizebox{3.6in}{!}{\includegraphics{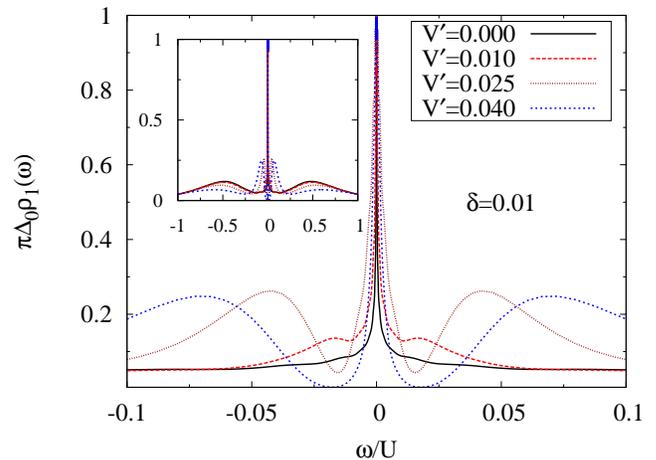}}}
\caption{\label{fig2}(color online)  Spectral function vs.\@ energy for
$V=0.1$, $\delta=0.01$ and   coupling $V^{\prime}=0.0$ (solid, black),
$V^\prime=0.01$ (dashed, red), $V^{\prime}=0.025$ (dotted, brown) and
$V^\prime=0.04$ (short-dash, blue).  The inset shows the same curves over
a wider energy range. Here, $U=0.5$, $U'=0$, and $D=1$.}
\end{figure}
We first fix $\delta$ ($=0.01$)
and vary the interdot coupling
$V^\prime$.  Figure~\ref{fig2}  shows $\rho_{1}(\omega)$ ($= \rho_{1\uparrow} = \rho_{1\downarrow}$
in Eq.\ \ref{rho1eq}),
for several values of $V^\prime$. The solid (black)
curve shows  the case $V^\prime=0$, which corresponds to the typical case of a
QD coupled to a conduction band with a flat density of states.  Notice the Kondo
resonance at the Fermi level in addition to the two Hubbard peaks at $\approx\pm U/2$ (see inset;
which give rise to Coulomb blockade peaks in the conductance at appropriate gate voltages).
The width of the Kondo peak is proportional to
the Kondo temperature, which in this case is \cite{T_K} $T_K^0 \equiv T_K(V'=0)=5.6\times
10^{-5}$. As $V^\prime$
increases, the low-energy structure of $\rho_1(\omega)$ is
strongly modified. Two satellite peaks emerge symmetrically about the central
peak at $\omega=0$. These peaks become more pronounced for larger $V^\prime$ and
the distance between them increases somewhat more than $2\sqrt{2{V^\prime}^2+ \delta^2}$, as one
would expect from the non-interacting regime.  The widths
of the satellite peaks increase as $V^\prime$ increases, similarly to the non-interacting case, where
their width is given by $\simeq \Delta_0 V'^2$.
Notice also Fano antiresonances in $\rho_1$ between the central and satellite peaks,
located at $\omega \approx \pm \delta$.  These antiresonances result from
destructive interference between the Kondo state of QD1 and the localized
levels in the lateral
QDs.\cite{torio} We will discuss this point in greater detail below.

\begin{figure}
\centerline{\resizebox{3.7in}{!}{
\includegraphics{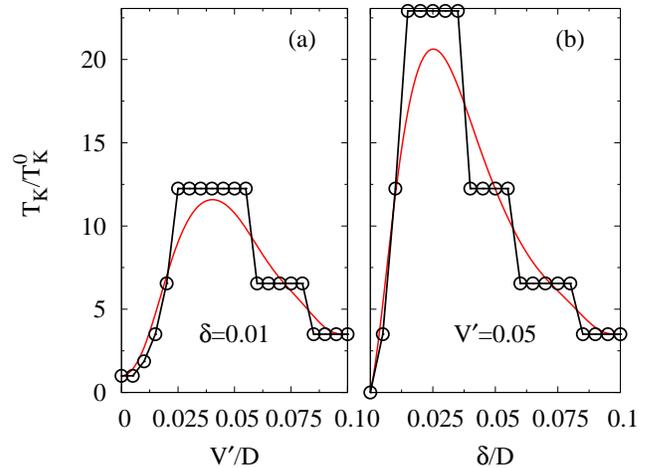}}}
\caption{\label{TK}(color online) (a) Kondo temperature as function of
$V^\prime$ for $\delta=0.01$.  Notice $T_K$ at $V^\prime=0$ is non-zero, $T_K^0=5.6\times10^{-5}$ here.
(b)  Kondo temperature as function of $\delta$ for $V^\prime=0.05$. 
Other parameters are as in Fig.\ \ref{fig2}.  Notice $T_K \rightarrow 0$ for $\delta \rightarrow 0$.  
The step structure seen here results from the NRG discretization of the
energy, and corresponds also to the discrete temperature values for which we calculate the
thermodynamics (see below). Continuous lines show smooth fit of
NRG results as guide to the eye.}
\end{figure}

It is interesting to notice that the central Kondo resonance peak width shows a
non-monotonic dependence on $V^\prime$, first increasing with $V^\prime$, but decreasing
after a particular value.  This behavior, first identified in Ref.\ \onlinecite{Trocha} from the
slave
boson approximation, suggests that the Kondo
temperature has a similar behavior, although this suspicion was not explicitly verified.
Using NRG, we find indeed that to be the case.  In Fig.\ \ref{TK}(a) we show
$T_K$ as function of $V^\prime$ for $\delta=0.01$,
and obtain a clearly convex curve with a maximum value ($\simeq 10T_K^0$) at $V^\prime \simeq 0.04$,
 in agreement
with the behavior seen in the Kondo resonance in Fig.\ \ref{fig2}.
We should emphasize that $T_K$ for $V^\prime > 0$ is \emph{larger} than for the isolated
QD1, $T_K^0$ (see value above).  

\begin{figure}
\centerline{\resizebox{3.6in}{!}{
\includegraphics{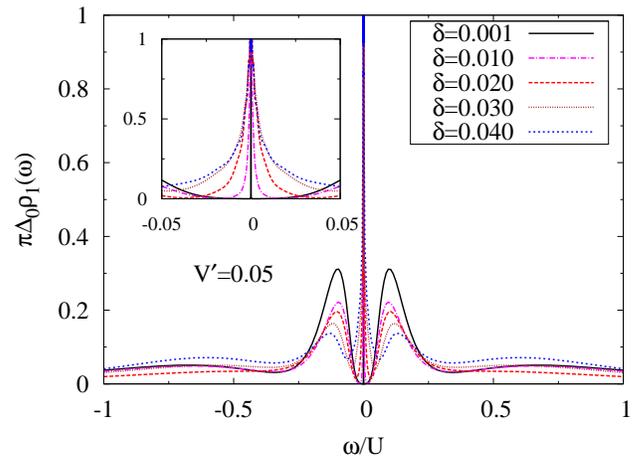}}}
\caption{\label{fig5}(color online)  Spectral function vs.\ energy for
$V=0.1$, $V^\prime=0.05$ and  $\delta=0.04$  (short-dash, blue), $\delta=0.03$
(dotted, brown),  $\delta=0.02$ (dashed, red),  $\delta=0.01$ (dot-dash,
magenta) and  $\delta=0.001$ (solid, black). The inset shows a zoom of the
region around $\omega=0$. }
\end{figure}

The zero-energy resonance also depends strongly on $\delta$.  Figure \ref{fig5} shows
$\rho_1(\omega)$ for $V^\prime=0.05$ and several values of $\delta$.  Notice
that all curves show well-defined peaks and their
positions remain nearly constant at either $\omega \approx 0$ or $\omega \approx \pm \sqrt{2{V^\prime}^2 + \delta^2}$.
However, the dips between the Kondo and the
satellite peaks are less pronounced for larger $\delta$, and
their position shifts with $\delta$, as can also be seen in the inset.
The displacement of the  dips toward the Fermi level as $\delta$ decreases
produces a ``squeezing" of the Kondo peak at the Fermi energy. The most evident
situation shown in Fig.\  \ref{fig5} is the case of $\delta=0.001$, when
the Kondo resonance takes the form of a narrow and sharp peak.
The behavior of $T_K$ with $\delta$ is shown in Fig.\  \ref{TK}(b), and it is clearly
non-monotonic as well. Notice that for
$V^\prime=0.$05 we find a maximum value  of $T_K \simeq 20T_K^0$ at $\delta\approx 0.025$.
More importantly, the
``squeezing'' of the Kondo resonance as $\delta\rightarrow 0$ is accompanied
by a strong suppression of $T_K$, which vanishes completely as $\delta \rightarrow 0$.
The above behavior is a clear signal of the influence of the subtunneling mode on the Kondo state.
Indeed, one can understand these results in terms of the Dicke state physics. 
For small values of $\delta$, the subtunneling Dicke state dominates
and the Kondo temperature decreases. For sufficiently large values of
$\delta$  the supertunneling Dicke state dominates and the Kondo
temperature reaches its maximum value. For larger  $\delta$ values, however, 
the system goes out of the Dicke regime and the Kondo temperature drops. \cite{Trocha}
We should also comment that the total suppression of the Kondo screening for $\delta =0$ is essentially a quantum phase
transition, which results in the QD cluster isolating itself from the current leads.  Perhaps
one way to view this suppression of Kondo screening is to think of a Kondo resonance in the central
dot that interferes destructively with the single-particle--like resonances of the side-connected dots (themselves hybridized with
the leads via the central dot).  The fact that all these resonances appear at the same energy, $\omega=0$, results in the 
eventual suppression of the Kondo state for the entire cluster.  Of course, this is only a qualitative argument, ultimately
validated by the NRG results.  The underlying physical mechanism for this destructive interference and resulting ground
state is fascinating, as we proceed to describe.

To provide a different perspective on the suppression  of the Kondo screening (and vanishing Kondo temperature),
we study the thermodynamical properties
of the system, namely the contribution of the impurity to the magnetic moment,
$\mu^2$, and the entropy, $S$, from which one extracts information
about the nature of the states the system visits when the parameters are
varied. As usual, $\mu^2(T)=\mu_{tot}^2(T)-\mu^2_0(T)$ and
$S(T)=S_{tot}(T)-S_0(T)$, where
$X_{tot}(T)$ and $X_0(T)$ (with $X=\mu^2,S$) are
thermodynamical quantities of the system calculated {\it with} and {\it
without}  impurity, respectively.
The magnetic moment and the entropy can be written in
terms of the magnetic susceptibility and the partition function as
$\mu^2(T)=T\chi(T)$, and $S=\log Z$ (we set
$k_B=1$).  In order to study the effects of all three dots, we employ here the ``cluster impurity"
approach described in Sec.\ \ref{CIA}.

\begin{figure}[h]
\centerline{\resizebox{3.6in}{!}{
\includegraphics{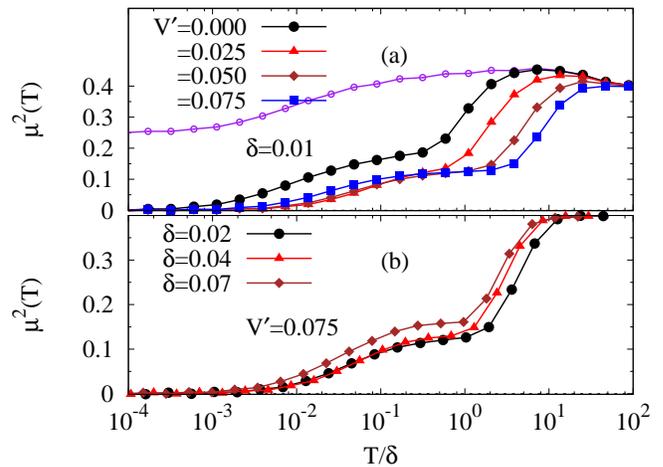}}}
\caption{\label{fig3}(color online) (a) Magnetic moment for the three-dot cluster as
function of  temperature for $\delta=0.01$  and
various values of $V^\prime$. Open small circles (purple) show results for $V^\prime = \delta=0$.
(b) Magnetic moment as  function of temperature
for $V^\prime=0.075$ and various values of  $\delta$. }
\end{figure}

\begin{figure}[h]
\centerline{\resizebox{3.6in}{!}{
\includegraphics{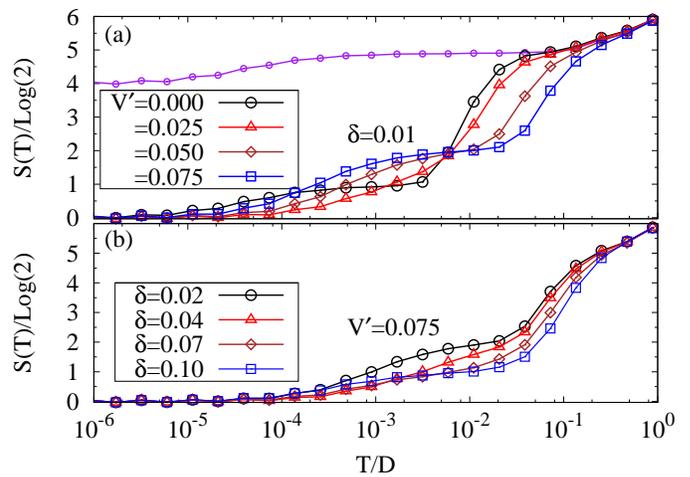}}}
\caption{\label{fig4}(color online)  (a) Entropy of three-dot cluster as function of
temperature for $\delta=0.01$ and  various values of $V^\prime$. Open small circles
(purple) show results for $V^\prime = \delta=0$. (b) Entropy of cluster as
function of temperature for $V^\prime=0.075$ and  various values of $\delta$. }
\end{figure}

Figures \ref{fig3}(a) and \ref{fig4}(a) show, respectively,  $\mu^2$ and $S$ as function of
the NRG iteration energy (temperature $T$) for several
values of $V^\prime$ and $\delta$.  Before engaging in the
discussion of the more complicated situations, lets us examine  the result for
$V^\prime=\delta =0$ (small open circles, purple),
corresponding to a single impurity case, as QD2 and QD3 have their level at the Fermi energy ($\delta=0$)
but are decoupled from the rest of the system ($V^\prime=0$)--we include this curve as a reference
for the more general cases.
For $T \gg U$, all the curves show convergence to
$\mu^2=3/8$ and $S=\log(64)$, resulting from the free orbital (FO) fixed point of
the system.  As $T\rightarrow 0$, DQ1 undergoes a crossover into a Kondo singlet state and
its magnetic moment is screened by the conduction band, while QD2 and QD3 remain in their
free orbital states (since $V^\prime =0$ and $\delta=0$), resulting in a net  unscreened
magnetic moment of $1/4$ and entropy $S=\log(16)$ for the three-dot cluster.

\begin{figure}[h]
\centerline{\resizebox{3.1in}{!}{
\includegraphics{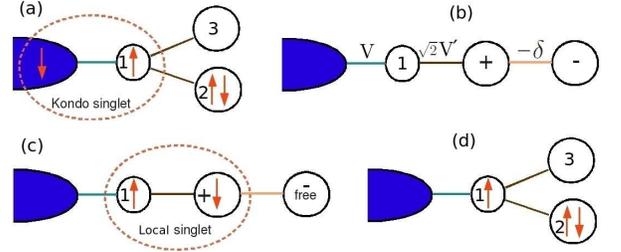}}}
\caption{\label{transf}(color online)  Schematic representation of the various
configuration of the systems. (a) Kondo singlet for $\delta\neq0$ and
$T<T_K$.
(b) Symmetric (``$+$'') and anti-symmetric
orbital (``$-$'') combination.     (c) Antiferromagnetic singlet formation between
$\varepsilon_1$ and ``$+$'' orbital.
(d) Magnetic moment
configuration for $\delta\neq0$ and
$T>T_K$.}
\end{figure}

For $V^\prime,\delta\neq0$,
different regimes can be found. In the low temperature limit ($T<T_K$) the system always
flows into a  Kondo state, where QD1 is in a Kondo singlet configuration with the
reservoirs, while QD2 and QD3 are combined into a symmetric/antisymmetric pair of states, one
of which is doubly occupied (and the other empty) at $T=0$, as schematically shown in Fig.\ \ref{transf}(a).
This situation is characterized by a magnetic
moment completely screened by the conduction band ($\mu^2=0$) and with entropy
$S=0$, as shown by all curves in Fig.\ \ref{fig3} and \ref{fig4}.

Other interesting situations occur when $T_K<T<\delta$. \cite{T_K} For $V^\prime \lesssim
\delta$, the physics can still be understood in terms of
the local orbitals, in which case  (since $T<\delta$)  QD2 and QD3 are still
locked in one doubly-occupied and one empty state.  In contrast, QD1 is in the magnetic
moment regime, contributing alone to the magnetic moment $\mu^2=1/8$ and to the
entropy $S=\log(2)$. When $V^\prime > \delta$,
however,  the system goes into a more intricate  configuration of local molecular orbitals.
This is better understood in terms of
symmetric (super-tunneling) and anti-symmetric (sub-tunneling) combination of
the QD2 and QD3 local orbitals,
defined as
\begin{eqnarray}
c_{+\sigma}&=&\frac{1}{\sqrt{2}}(c_{2\sigma}+c_{3\sigma})\nonumber \\
c_{-\sigma}&=&\frac{1}{\sqrt{2}}(c_{2\sigma}-c_{3\sigma}),
\end{eqnarray}
where $c_{+\sigma}$  and  $c_{-\sigma}$ annihilate an electron in the bonding
and anti-bonding orbital, respectively, with energy
$\varepsilon_+=\varepsilon_-=0$ (for $\varepsilon_3=-\varepsilon_2=\delta$ as before).
With this transformation, the Hamiltonian of the dots reads
\begin{eqnarray}
\label{H_dotst}
\tilde
H_{dots}&=&\sum_{\sigma}\varepsilon_1c^\dagger_{1\sigma}c_{1\sigma} +Un_{
1\uparrow}n_{1\downarrow}\nonumber\\
&&+\sqrt{2}V^\prime\sum_{\sigma}\left[c^\dagger_{1\sigma}c_{+\sigma}
+h.c.\right]\nonumber\\
&&-\delta \sum_{\sigma}\left[c^\dagger_{+\sigma}c_{-\sigma}
+h.c.\right].
\end{eqnarray}
The transformed system is schematically represented in Fig.\ \ref{transf}(b).
Note that the
orbital ``$+$'' and ``$-$'' are coupled to each other by a matrix element
$-\delta$ and only the orbital ``$+$" couples to the QD1 via matrix
element $\sqrt{2}V^\prime$. On this transformed basis we can provide a
more intuitive analysis of the thermodynamic quantities  as follows.  For $T_K<T<\delta$
and $V^\prime \gg \delta$ the configuration of the system is represented in
Fig.\ \ref{transf}(c), where the
QD1 orbital couples to the ``$+$'' orbital to form a ``local singlet" (essentially
decoupled from the leads), while the
``$-$'' orbital remains almost free (since $\delta\ll V^\prime$).
The contribution to the magnetic moment and entropy is then provided
solely by the ``$-$'' orbital,
resulting in $\mu^2=1/8$ and $S=\log(4)$, as clearly seen in the
curves with (blue) square symbols in Fig.\ \ref{fig3}(a) and \ref{fig4}(a).  We should notice
that the two curves for $V^\prime=0$ (for $\delta=0$ and $\delta=0.01$)--in each of the (a) panels of both figures--are identical
for temperatures $T \gg \delta=0.01$, since for high $T$ the QD2 and QD3 levels at $\pm \delta$ are thermally
accessible (which is no longer the case when $T \ll \delta$). 

In Fig.\ \ref{fig3}(b) and
 \ref{fig4}(b) we fix $V^\prime=0.075$ and plot the magnetic moment and
entropy for different values of $\delta$. For $\delta=0.02\ll V^\prime$,
the situation described above is favored, but when $\delta\rightarrow V^\prime$,
the plateaus in the
susceptibility and entropy shift towards the
curves with empty (black) circles of panel (a) in the respective figures.  The system
is then returning to
the situation when the local orbitals are described independently, as shown in
Fig.\ \ref{transf}(d).  Notice the Kondo singlet is not yet formed at these high temperatures
($T > T_K$).
These results reveal the establishment of a local
antiferromagneticaly correlated state, involving orbitals $\varepsilon_1$ and
``$+$'', emerging at temperatures above $T_K$.

\begin{figure}[h]
\centerline{\resizebox{3.5in}{!}{
\includegraphics{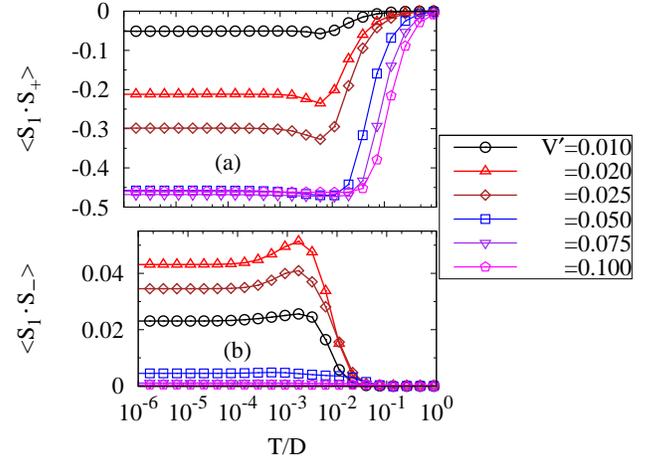}}}
\caption{\label{correlation}(color online) Spin-spin correlations as function of
$T$ for $\delta=0.01$ and various values of $V^\prime$.  Lower panel proves
that the spin correlation of QD1 with ``$-$'' is vanishingly small.  In contrast, upper panel
exhibits strong antiferromagnetic correlation with the ``$+$'' orbital with increasing $V^\prime$.}
\end{figure}

In order to provide more evidence  of the local
singlet between QD1 and the ``$+$'' orbital, Fig.\ \ref{correlation} depicts the spin-spin correlation
value $\langle {\bf S_1}\cdot {\bf S_+}\rangle$ (top panel); the correlation with the ``$-$''
orbital, $\langle {\bf
S_1}\cdot {\bf S_-}\rangle$ is shown in the bottom panel, as function of temperature for
various values of $V^\prime$. Notice that $\langle {\bf S_1}\cdot
{\bf S_+}\rangle$  is always negative (antiferromagnetic) and has increasingly negative values for
larger
$V^\prime$.  In contrast, the correlation between the spin in
QD1 and in the ``$-$'' orbital is an order of magnitude weaker, and totally vanishes for  large
$V^\prime$.
It is interesting to notice that although the local correlation with the ``$+$'' orbital is large,
it does not fully
exhaust the anticipated singlet correlation of $-3/4$, even for the larger $V^\prime$ values, where
it seems
to saturate at $\simeq -1/2$.

\subsection{Effects of particle-hole asymmetry} \label{phasym}

In a typical experimental situation, asymmetry in the structure parameters is more likely to
occur. It is therefore of interest to analyze the stability and behavior of the results discussed above
with respect to changes that more generally make the problem particle-hole asymmetric.  
We introduce asymmetry
parameters $\kappa_v$ and $\kappa_\delta$ that allow for $V_{12}\neq V_{13}$, and
$|\epsilon_3|\neq|\epsilon_2|$, respectively as:
\begin{eqnarray}
 V_{1l}=V^\prime(1\pm\kappa_v)
\end{eqnarray}
and
\begin{eqnarray}
 \epsilon_{l}=\delta(1\pm\kappa_\delta),
\end{eqnarray}
where the positive (negative)  sign  in the expressions above corresponds to the index $l=2~(3)$. The
fully symmetric situation previously described corresponds obviously to the case
$\kappa_v=\kappa_\delta=0$. 

\begin{figure}
\centerline{\resizebox{3.4in}{!}{
\includegraphics[angle=0]{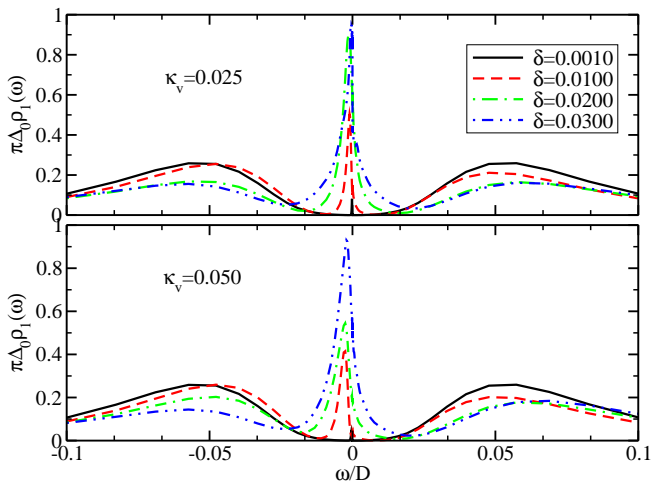}}}
\caption{\label{fig9}(color online)  Spectral function vs.\ energy for asymmetric coupling to the
side dots. Top panel ($\kappa_v=0.025$)  and bottom panel ($\kappa_\delta=0.05$)
$V=0.1$, $V^\prime=0.05$ and  $\delta=0.001$ (solid, black), $\delta=0.01$
(dashed, red), $\delta=0.02$ (dot-dash, green), and $\delta=0.03$ (dot-dot-dash, blue).}
\end{figure}

We first allow for asymmetry in the couplings to the side dots;
Fig.\ \ref{fig9} shows the results for weakly asymmetric couplings $V_{12}$ and $V_{13}$ for different
values of $\delta$. The top (bottom) panel shows the case of $\kappa_v=0.025$ ($\kappa_v=0.05$).
The asymmetry in the couplings results in a shift of the central peak away
from the Fermi level, as well as its strong suppression (compare with Fig.\ \ref{fig5}). 
This effect is less pronounced for larger values of $\delta$.
Notice, for instance, that for $\delta=0.001$ the central peak is almost completely suppressed. Also,
comparing the curves in the top and bottom panels, we observe that for a given $\delta$ the shift
and suppression of the peak is larger as $\kappa_v$ increases.

\begin{figure}
\centerline{\resizebox{3.4in}{!}{
\includegraphics[angle=0]{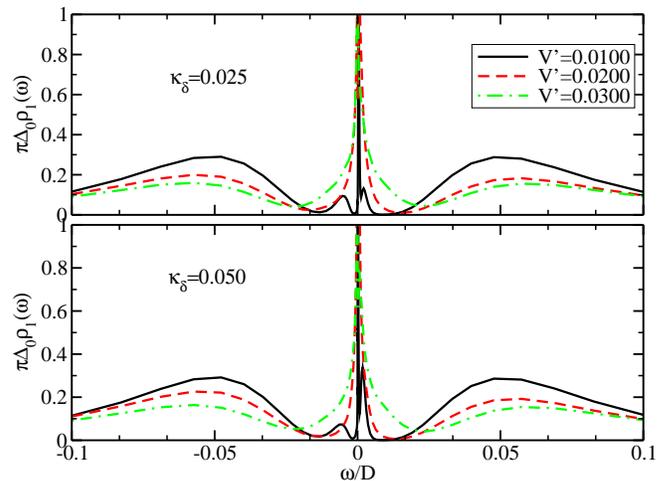}}}
\caption{\label{fig10}(color online)  Spectral function vs.\ energy for the asymmetric bare levels about the Fermi level, with 
asymmetries $\kappa_\delta=0.025$ (top) and $\kappa_\delta=0.050$ (bottom). The other parameters are $V=0.1$, $\delta=0.01$ 
and  $V^\prime=0.01$  (solid, black), $V^\prime=0.02$ (dashed, red), and $V^\prime=0.03$ (dot-dash, green). }
\end{figure}

In Fig.\ \ref{fig10}  we fix $\delta=0.01$ and $\kappa_v=0$ (symmetric couplings to central dot),
and show the spectral function as
function of energy for two values of $\kappa_\delta$ and different values of $V^\prime$. In the top
(bottom) panel we have $\kappa_\delta=0.025$ ($\kappa_\delta=0.05$). We notice that for small
$V^\prime$ (for instance $V^\prime=0.01$), there is a strong
distortion in the spectral function near the Fermi level for both values of $\kappa_\delta$. 
However this distortion is much weaker when $V^\prime$ increases, so that for $V^\prime=0.02$ and $0.03$, the asymmetry in the spectral function is nearly absent.

\begin{figure}
\centerline{\resizebox{3.4in}{!}{
\includegraphics[angle=0]{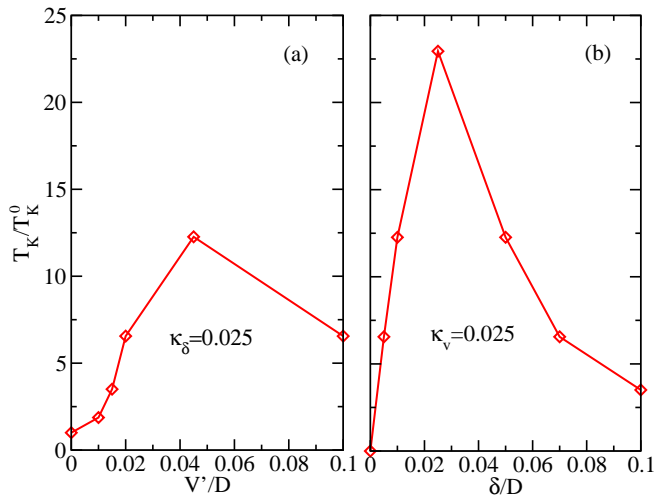}}}
\caption{\label{TKSA}(color online) Kondo temperature for asymmetric three-dot systems.  Panel (a) shows $T_K$ (in units of $T_K^0$ -- see caption Fig.\ \ref{TK}) for system of Fig.\ \ref{fig10}a.  Panel (b) shows $T_K$ for system of Fig.\ \ref{fig9}a.  Notice strong enhancement of $T_K$ in both cases with increasing parameter.}
\end{figure}

One would expect, from seeing these results for the spectral functions, that the particle-hole asymmetry in the system would introduce non-monotonic changes in the Kondo temperature.  In Fig.\ \ref{TKSA} we illustrate this point.  Panel (b) shows the associated Kondo temperature for the system parameters of Fig.\ \ref{fig9}a, where the coupling to the side dots is asymmetric (with a value of $\kappa_v=0.025$).  As suggested by the increasing width of the zero-energy peak in Fig.\ \ref{fig9}a, $T_K$ increases with $\delta$, reestablishing the Kondo state that was absent in the full Dicke regime (for $\delta =0$), while  larger values eventually reduce $T_K$, although {\em not to zero}.  Similarly, the asymmetry in side dot level location illustrated in Fig.\ \ref{fig10}a, results in a rapid increase of $T_K$ with $V'$, shown in panel (a), so that the system has an order of magnitude larger Kondo temperature for $V' \approx 0.05$, before dropping for larger $V'$ values. 

\begin{figure}
\centerline{\resizebox{3.4in}{!}{
\includegraphics[angle=0]{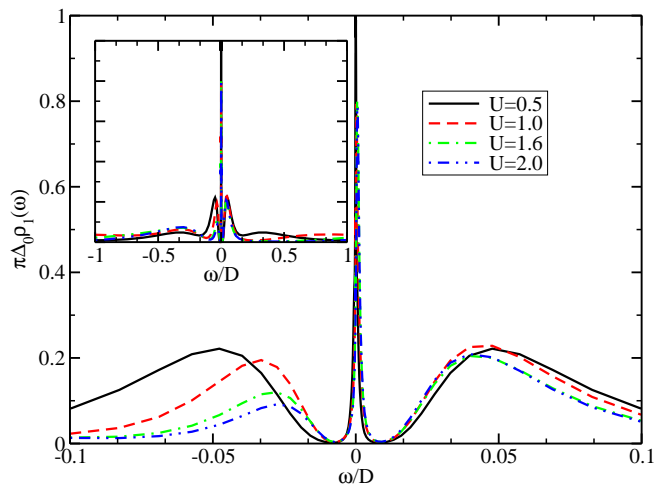}}}
\caption{\label{fig11}(color online)  Spectral function vs.\ energy for
$V=0.1$, $V^\prime=0.05$,  $\delta=0.01$ and $U=0.5$ (solid, black), $U=1.0$
(dashed, red),  $U=1.6$ (dot-dash, green),  and $U=2.0$ (dot-dot-dash,
blue) (solid, black). The inset shows  the same curves for the entire
region within the conduction band.}
\end{figure}

An additional important source of particle-hole asymmetry is the bare level position of QD1. In
order to explore how it affects our results, we allow the level $\epsilon_1$ to be different from
$-U/2$. For convenience, we keep $\epsilon_1=-0.25$ and change $U$, which allows not
only to explore the effect of the particle-hole asymmetry but also the effect of larger values of $U$.
Figure \ref{fig11} shows the spectral function vs.\ energy for
$\delta=0.01$ and various values of $U$. The curve for $U=0.5$, corresponding to the symmetric case
studied in Sec.\ \ref{symmetric}, exhibits two peaks symmetrically placed about
the Fermi level, as well as Hubbard bands located near $\epsilon_1$ and $\epsilon_1+U$, as can be
seen in the inset of the figure. As $U$ increases, one clearly observes that the curves no longer
display particle-hole symmetry. We observe a decreasing height of the central peak, as well as a minor shift
in position, while its width remains almost unchanged with increasing $U$. More interesting, we notice that
the satellite peak below the Fermi level is suppressed and shifted toward $\omega=0$, while the
corresponding peak on the positive side is nearly unchanged. This is a clear consequence of breaking 
particle-hole symmetry in the problem.

\begin{figure}
\centerline{\resizebox{3.4in}{!}{
\includegraphics[angle=0]{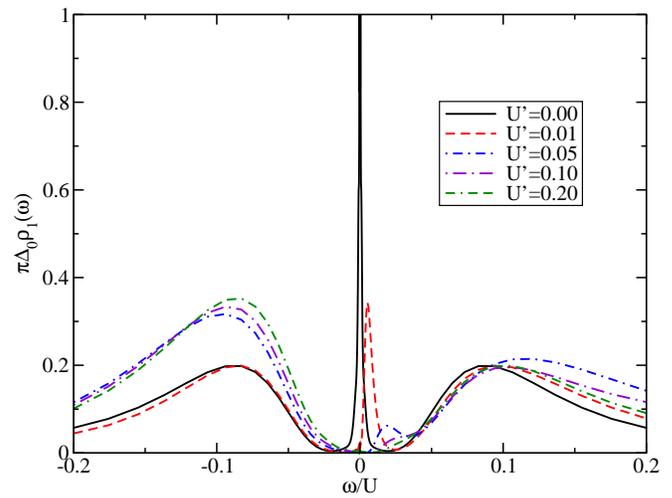}}}
\caption{\label{Uprime}(color online) Spectral function for the three dot cluster with non-zero Coulomb interaction in all dots for energies close to the Fermi level.  The solid (black) curve with $U'=0$ represents the same system as in Fig.\ \ref{fig2}, $\delta=0.01$, $V'=0.05$, $V=0.1$, and $U=0.5$.  As $U'$ increases, the asymmetry is evident, strongly affecting the Kondo peak near zero energy. }
\end{figure}

We have also investigated the role of non-vanishing Coulomb interactions in the side dots QD2 and QD3.  
To that effect, the problem is addressed using NRG for the three quantum dot complex treated as a cluster impurity, 
as described in Sec.\ \ref{CIA}, and with the inclusion of a non-zero Coulomb interaction $U_2=U_3=U'$.  
Figure \ref{Uprime} presents the spectral function for different values of $U'$, for QD2 and QD3 levels placed asymmetrically about the Fermi level ($\delta=0.01$).  For $U'=0$ the system is particle-hole symmetric (as $\varepsilon_1=-U/2$), as also shown in Fig.\ \ref{fig2}.  For increasing $U'$, however, the asymmetry in the spectral function is evident, strongly shifting and suppressing the Kondo resonance near the Fermi level.   This indicates the important role that Coulomb interactions in the nearby dots have in the overall correlations present in the system, and suggests that the observation of the Dicke effect is indeed a delicate undertaking.

\subsection{Conductance}
We now turn our attention to the effect on the conductance of the system of the various
features discussed above.
It is clear that the details of the density of states at the Fermi level are directly connected to the
zero-temperature transport properties in linear response.
In the zero-bias limit, the conductance of the system
can be obtained by the Landauer formula,
$G=(2e^2/h)\int_{-\infty}^\infty -{\tt Im}[{\cal T}(\omega)]\left(
\partial f(\omega)/\partial \omega\right)d \omega$,
where $f(\omega)$ is the Fermi distribution function and, in the present  case,
${\cal T}(\omega) =\Delta_0G_{dd}(\omega)$. At $T=0$ the
conductance reduces to
$G=(2e^2/h)\pi\Delta_0\rho_{1}(E_F)$.
From the density of states in Fig.\ \ref{fig2}
and \ref{fig5} we observe that for all values of $V^\prime$ and $\delta>0$,
the DOS always exhibits a Kondo peak at the Fermi level (which
becomes very narrow and sharp as $\delta\rightarrow 0$).  Since the
conductance of the system is proportional to the density
of states of DQ1, this results in unitary conductance ($2e^2/h$) for all values of
$V^\prime$ and $\delta>0$, discontinuously dropping to zero at $\delta=0$.
The above behavior is explained by the complete localization of the subtunneling state for $\delta=0$.
In this case the state has no projection on the central quantum dot and therefore it does not contribute to the conduction. 
Similarly, one would expect vanishing conductance for small bias as long as $\delta=0$. \cite{Trocha}

\begin{figure}
\centerline{\resizebox{3.4in}{!}{
\includegraphics[angle=0]{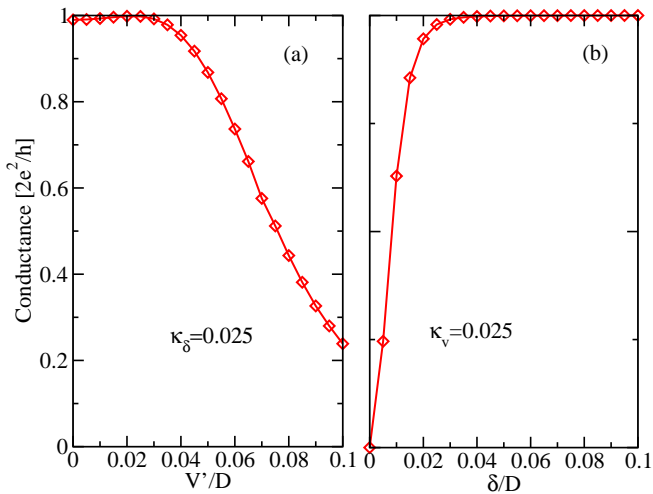}}}
\caption{\label{figlast}(color online) Conductance for asymmetric systems corresponding to Fig.\ \ref{TKSA}, for each of the two panels there. } 
\end{figure}

An interesting result of the particle-hole asymmetry, as we have seen in the previous section, is to shift the Kondo resonance away from the Fermi level, as well as to change its amplitude.  This has direct impact on the linear conductance of the structure.  Figure \ref{figlast} illustrates this behavior for different asymmetries (notice in fact that although $\kappa_v$ and $\kappa_\delta$ are constant in this figure, increasing $V'$ or $\delta$ value increases the asymmetry of the structure).  In both cases, the conductance is indeed found to change drastically with the appropriate parameter.  It is interesting that asymmetry in the QD2 and QD3 level position (for $\kappa_\delta=0.025$) and increasing dot coupling $V'$ results in a strong suppression of the linear conductance, even as the Kondo temperature of the system (Fig.\ \ref{TKSA}a) is {\em increasing}.  On the other hand, the enhancement of $T_K$ that we see for the coupling asymmetry ($\kappa_v=0.025$) in Fig.\ \ref{TKSA}b, does produce a rapid increase in conductance for larger $\delta$ values.

\section{Conclusions}
In summary, we have investigated the interplay of Dicke and Kondo effects in a central strongly interacting
quantum dot coupled to a conduction band and two localized levels provided by additional
nearly non-interacting dots. We study in detail the Kondo regime of the system, by
applying
a numerical renormalization group analysis to a finite-$U$  multi-impurity Anderson
Hamiltonian model. We find that the system displays
a ``squeezed"  Kondo resonance  as we tune the single particle levels of  the
side-coupled dots toward the Fermi level.  In the strict Dicke limit of degenerate
single-particle levels, the Kondo singlet disappears, as its Kondo temperature
vanishes.
This behavior is found to be closely related to the formation of  a local singlet
state involving  the (coupled) orbitals of the interacting
dot and the super-tunneling state (the  symmetric combination of the single
particle levels of the two side coupled dots). This local singlet results in the suppression
of the Kondo screening, which is detrimental to the conductance of the system.
We have further explored the consequences of system asymmetries, introduced by different couplings, 
level locations and Coulomb interactions.  These are found to strongly affect the spectral density and 
the conductance through the structure.  We believe that our study not only
clarifies the relevant picture of an interacting system coupled to side quantum dots and conduction band,  but it also provides important guidelines for experimental realization  of these ideas.

\section{Acknowledgements}
We thank helpful discussions with N. Sandler and G. B. Martins and financial support from
CNPq, CAPES, FAPEMIG, NSF-PIRE, and MWN/CIAM (CNPq, CONICYT, NSF-DMR). P.A.O acknowledges  financial
support from CONICYT/Programa
Bicentenario de Ciencia y Tecnologia (CENAVA, grant ACT27) and FONDECYT, under
grant 1080660.


\begin{thebibliography}{apssamp}
\bibitem{DGG} %Kondo effect in a single-electron transistor
D. Goldhaber-Gordon, H. Shtrikman, D. Mahalu, D. Abusch-Magder, U. Meirav, and M. A. Kastner,
Nature {\bf 391}, 156 (1998).

\bibitem{Inoshita} T.~Inoshita, Science {\bf 281}, 526 (1998).

\bibitem{Cronenwett} S.~M.~Cronenwett, T.~H.~Oosterkamp, and L.~P.~Kouwenhoven,
Science {\bf 281},  540 (1998).

\bibitem{Sasaki} S.~Sasaki {\it et al.}, Nature {\bf 405}, 764 (2000).

\bibitem{Pustilnik} M.~Putilsnik and L.~I.~Glazman, Phys. Rev. Lett. {\bf 87},
216601 (2001)

% double quantum dot
\bibitem{busser}  C.~A.~B\"usser {\it et al}, Phys. Rev. B {\bf 62}, 9907 (2000).

\bibitem{jeong}  H.~Jeong, A.~M.~Chang, M.~.R.~Melloch, Science {\bf 293}, 2221 (2001).

\bibitem{cornaglia}  P.~S.~Cornaglia, D.~R.~Grempel, Phys. Rev. B {\bf 71},
075305 (2005).

\bibitem{vernek} E.~Vernek, N. Sandler, S. E. Ulloa, and E. V. Anda, Phys. E {\bf 34}, 608 (2006).

\bibitem{8l} L.~G.~G.~V.~D.~Dias da Silva, N.~Sandler, K.~Ingersent, and
S.~E.~Ulloa, Phys. Rev. Lett. {\bf 97}, 096603 (2006); L.~G.~G.~V.~D.~Dias da
Silva, N.~Sandler, P. Simon K.~Ingersent, and
S.~E.~Ulloa, Phys. Rev. Lett. {\bf 102}, 166806 (2009).

\bibitem{sasaki3} S.~Sasaki, H.~Tamura, T.~Akazaki, and T.~Fujisawa, Phys. Rev Lett. {\bf 103},
266806 (2009).

\bibitem{zitko5} R.~\v{Z}itko, arXiv:1001.1983v2 [cond-mat.mes-hall].

% Triple quantum dots coupled  in series, infinite-U slave boson approach
\bibitem{kuzmenko1}  T.~Kuzmenko, K.~Kikoin, Y.~Avishai, Europhys. Lett. {\bf 64}, 218
(2003).

\bibitem{jiang} Z.~T.~Jiang, Q. F. Sun, and Y. P. Wang, Phys. Rev. B {\bf 72},
045332 (2005).


\bibitem{zitko2} R.~\v{Z}itko and J. Bon\v{c}a, A. Ram\v{s}ak, and T. Rejec,
Phys. Rev. B {\bf 73}, 153307 (2006).

\bibitem{lobos} A. M. Lobos,  and A. A. Aligia, Phys. Rev. B {\bf 74},
165417 (2006).

\bibitem{Vernek1}  E.~Vernek, C.~A.~B\"usser, G.~B.~Martins,
E.~V.~Anda, N.~Sandler, and S.~E. Ulloa, Phys. Rev. B {\bf 80}, 035119
(2009).

\bibitem{Numata} T. Numata, Y. Nisikawa, A. Oguri, and A. C. Hewson,
Phys. Rev. B {\bf 80}, 155330 (2009).

\bibitem{Chiappe1} G.~Chiappe, E.~V.~Anda, L.~Costa Ribeiro, and E.~Louis,
Phys. Rev. B {\bf 81}, 041310 (2010).


\bibitem{klaus}  R. Leturcq, L. Schmid, K. Ensslin, Y. Meir, D. C. Driscoll, and A. C. Gossard,
Phys. Rev. Lett. {\bf 95}, 126603 (2005).
%Probing the Kondo Density of States in a Three-Terminal Quantum Ring

\bibitem{haug} M.~C.~Rogge and  R.~J.~Haug Phys. Rev. B {\bf 77}, 193306
(2008).

\bibitem{fano} U. Fano, Phys. Rev. \textbf{124}, 1866 (1961).

\bibitem{Hofstetter} W.~Hofstetter, J.~Konig, and H.~Schoeller,
Phys. Rev. Lett. {\bf 87}, 156803 (2001).

\bibitem{17'} %Kondo screening suppression by spin-orbit interaction in quantum dots
E. Vernek, N. Sandler, and S. E. Ulloa, Phys. Rev. B {\bf 80}, 041302(R) (2009).

\bibitem{sato} M. Sato, H. Aikawa, K. Kobayashi, S. Katsumoto, and Y. Iye, Phys.
Rev. Lett. {\textbf{95}}, 066801 (2005).

\bibitem{WF} D. Withoff and E. Fradkin, Phys. Rev. Lett. {\bf 64}, 1835 (1990).

\bibitem{Ingersent} K. Ingersent, Phys. Rev. B {\bf 54}, 11936 (1996); C. Gonzalez-Buxton and K.
Ingersent,  Phys. Rev. B {\bf 57}, 14254 (1998).

\bibitem{orellana} P. A. Orellana, G.A. Lara, and E. V. Anda, Phys. Rev. B
{\textbf{74}}, 193315 (2006).

\bibitem{dicke} R. H. Dicke, Phys. Rev. {\textbf 89}, 472 (1953).

\bibitem{18S} T.~V.~Shahbazyan and S.~E.~Ulloa, Phys. Rev. Lett. {\bf 79}, 3478(1997); Phys. Rev. B
{\bf 57}, 6642
(1998).

\bibitem{18C} A.~L.~Chudnovskiy, and S.~E.~Ulloa, Phys. Rev. B {\bf 63}, 165316
(2001).

\bibitem {Brandes} T. Brandes, Phys. Rep. \textbf{408}, 315 (2005).

\bibitem{Debray} P. Debray, O. E. Raichev, P. Vasilopoulos, M. Rahman, R.
Perrin, W. C. Mitchell, Phys. Rev. B {\textbf{6}1} 10950 (2000).

\bibitem{torio}  M.~E.~Torio, K.~Hallberg, A.~H.~Ceccatto, C.~R.~Proetto, Phys.
Rev. B {\textbf{65}}, 085302 (2002).

\bibitem{Trocha} P.~Trocha and J.~Barna\'s, Phys. Rev. B {\textbf{78}}, 075424
(2008).

\bibitem{Wilson} K.~Wilson, Rev. Mod. Phys.  {\bf 47} 773 (1975).

\bibitem{Murthy} H.~R.~Krishna-murthy, J.~W.~Wilkins, and K.~Wilson, Phys. Rev.
B {\bf 21}, 1003 (1980);
 H.~R.~Krishna-murthy, J.~W.~Wilkins, and K.~Wilson, Phys. Rev. B {\bf 21}, 1044
(1980).

\bibitem{24B} R. Bulla, T. A. Costi, and T. Pruschke,
%Numerical renormalization group method for quantum impurity systems
Rev. Mod. Phys. {\bf 80}, 395 (2008).

 \bibitem{Jayaprakash} K.~Chen and C.~Jayaprakash, Phys. Rev. B {\bf 20}, 14436
(1995).

 \bibitem{Fetter} L.~A. Fetter and J. D. Walecka, {\it Quantum Theory
of Many-Particle Systems} (Dover, New York, 2003).


\bibitem{Bulla1} R.~Bulla, T.~A.~Costi, and D.~Vollhardt, Phys. Rev. B {\bf
64}, 045103 (2001).

\bibitem{Kramers} ${\tt
Re}[G(\omega)]=\int_{-\infty}^\infty(\omega-\omega^\prime)^{-1}{\tt
Im}[G (\omega^\prime)]\,d\omega^\prime$. We typically keep 1200 states in the NRG runs and use
a discretization parameter $\Lambda=2.5$.

\bibitem{T_K} The Kondo temperature is obtained using Wilson's criterion,
{\it e.g.}, $\mu^2(T_K)=0.0701$.

 \end{thebibliography}
\end{document}